\documentclass{article}
\usepackage[dvips]{color}
\usepackage{amsmath}
\usepackage{amsfonts}   
\usepackage{amssymb}    
\usepackage{multicol}
\usepackage{mathptm,graphicx}
\usepackage{epsf}
\newtheorem{theorem}{Theorem}
\newtheorem{lemma}{Lemma}
\newtheorem{remark}{Remark}
\newtheorem{definition}{Definition}

\newtheorem{assum}{Assumption}

\begin{document}
\title{Fractional G-White Noise Theory, Wavelet Decomposition for Fractional
G-Brownian Motion, and Bid-Ask Pricing Application to Finance
Under Uncertainty}
\author{Wei Chen \\
Institute of Quantitative Economics\\
 School of Economics\\
    Shandong University\\
    250100, Jinan, China\\
weichen@sdu.edu.cn }
\date{}
\maketitle
\begin{center}
\begin{minipage}{120mm}
\baselineskip 0.2in {\small {\bf Abstract} G-framework is
presented by Peng \cite{PengC} for measure risk under uncertainty.
In this paper, we define fractional G-Brownian motion (fGBm).
Fractional G-Brownian motion is a centered G-Gaussian process with
zero mean and stationary increments in the sense of sub-linearity
with Hurst index $H\in (0,1)$. This process has stationary
increments, self-similarity, and long rang dependence properties
in the sense of sub-linearity. These properties make the
fractional G-Brownian motion a suitable driven process in
mathematical finance. We construct wavelet decomposition of the
fGBm by wavelet with compactly support. We develop fractional
G-white noise theory, define G-It\^o-Wick stochastic integral,
establish the fractional G-It\^o formula and the fractional
G-Clark-Ocone formula, and derive the G-Girsanov's Theorem. For
application the G-white noise theory, we consider the financial
market modelled by G-Wick-It\^o type of SDE driven by fGBm. The
financial asset price modelled by fGBm has volatility uncertainty,
using G-Girsanov's Theorem and G-Clark-Ocone Theorem, we derive
that sublinear expectation of the discounted European contingent
claim is the bid-ask price of the claim.
}
{\small \bf MSC-claasification: 60E05, 60H40,60K,60G18,60G22}\\
{\small \bf JEL-claasification: G10,G12,G13}
\end{minipage}
\end{center}
\newpage
\section{Introduction}

The stochastic process $B_H^{0}(t)$, which is continuous Gaussian
process with stationary increments
\begin{eqnarray*}
E[B^0_H(t)B^0_H(s)]=\frac{1}{2}[|t|^{2H}+|s|^{2H}-|t-s|^{2H}],\ \
H\in (0,1),\ \ (E[\cdot]\ \ \mbox{is some linear expectation})
\end{eqnarray*}
was originally introduced by Kolmogorov \cite{Kol40} (1940) in
study of turbulence under the name "Wiener Spiral", and the
process has the self-similarity property: for $a>0$
\begin{eqnarray*}
Law(B^0_H(at),\ t\ge 0)=Law(a^HB^0_H(t),\ \ t\ge 0).
\end{eqnarray*}
Later, when the papers of Hurst \cite{Hur51} (1951) and Hurst,
Black and Simaika \cite{HBS65} (1965) devoted to long-term storage
capacity in Nile river, were published, the parameter $H$ got the
name "Hurst parameter". The current name fractional Brownian
motion (fBm) comes from the other pioneering paper by Mandelbrot
and Van Ness \cite{MvN68} (1968), in which the stochastic calculus
with respect to the fBm was considered. The fractional Brownian
motion has similarity property and long rang dependence property,
which was leaded to describe a great variety of natural and
physical phenomena, such as, hydrodynamics, natural images,
traffic modelling in broadband networks, telecommunications, and
fluctuations of the stock market.

The first continuous-time stochastic model for a financial asset
appeared in the thesis of Bachelier \cite{Bachelier} (1900). He
proposed modelling the price of a stock with Brownian motion plus
a linear drift.The drawbacks of this model are that the asset
price could become negative and the relative returns are lower for
higher stock prices. Samuelson \cite{Samuelson} (1965) introduced
the more realistic model
\begin{eqnarray*}
S_t=S_0\exp{((\mu-\frac{\sigma^2}{2})+\sigma B_t^0)},
\end{eqnarray*}
which have been the foundation of financial engineering. Black and
Scholes \cite{Black} (1973) derived an explicit formula for the
price of a European call option by using the Samuelson model with
$S_0=\exp{(rt)}$ through the continuous replicate trade. Such
models exploded in popularity because of the successful option
pricing theory, as well as the simplicity of the solution of
associated optimal investment problems given by Merton
\cite{Merton} (1973).

However, the Samuelson model also has deficiencies and up to now
there have been many efforts to build better models. Cutland et
al. \cite{Cutland} (1995) discuss the empirical evidence that
suggests that long-range dependence should be accounted for when
modelling stock price movements and present a fractional version
of the Samuelson model. For $H\in (\frac{1}{2},1)$ the fractional
Gaussain noise $B_H^0(k+1)-B_H^0(k)$ exhibits long-range
dependence, which is also called the Joseph effect in mandelbrot's
terminology \cite{Mandelbrot} (1997), for $H=\frac{1}{2}$ the fBm
is semimartingale and all correlations at non-zero lags are zero,
and for $H\in (0,\frac{1}{2})$ the correlations sum up to zero
which is less interesting for financial applications
\cite{Cutland} (1995). However, empirical evidence is given of a
Hurst parameter with values in $(0,\frac{1}{2})$ for foreign
exchange rates \cite{Los}.

Hu and $\O$ksendal \cite{Hu} (2003) develop fractional white noise
theory in a white noise probability space
$(S^{\prime}(R),\mathcal{F})$ with $\mathcal{F}$ the Borel field,
modelled the financial market by Wick-It\^o type of stochastic
differential equations driven by fractional Brownian motions with
Hurst index $H\in (\frac{1}{2},1)$, and compute explicit the price
and replicating portfolio of a European option in this market.
Elliott and Hoek \cite{Elliott} (2003) present an extension of the
work of Hu and $\O$ksendal \cite{Hu} for fractional Brownian
motion in which processes with all indices include $H\in (0,1)$
under the same probability measure, describe the financial market
by a SDE driven by a sum of fractional Brownian motion with
various Hurst indices and develop the European option pricing in
such a market.

In an uncertainty financial market, the uncertainty of the
fluctuation of the asset price comes from the drift uncertainty
and the volatility uncertainty. For the drift uncertainty, in the
probability framework Chen and Epstein \cite{Chen} (2002) propose
to use g-expectation introduced by Peng in \cite{Peng1} (1997) for
a robust valuation of stochastic utility. Karoui, Peng and Quenez
in \cite{PengKaroui} and Peng in \cite{Peng} (1997) propose to use
time consistent condition g-expectation defined by the solution of
a BSDE, as bid-ask dynamic pricing mechanism for the European
contingent claim. Delbaen, Peng and Gianin (\cite{DelbaenPeng})
(2010) prove that any coherent and time consistent risk measure
absolutely continuous with respect to the reference probability
can be approximated by a g-expectation.

In the probability framework, the volatility uncertainty model was
initially studied by Avellaneda, Levy and Paras \cite{Avell}
(1995) and Lyons \cite{Lyons} (1995) in the risk neutral
probability measures, they intuitively give the bid-ask prices of
the European contingent claims as superior and inferior
expectations corresponding with a family of equivalent
probability.

There is uncertainty in economics, and no one knows its
probability distribution. Almost all the financial market
fluctuations show volatility uncertainty (VIX, S$\&$P 500, Nasdaq,
Dow Jones, Eurodollar, and DAX, etc), and the volatility
uncertainty is the most important, interesting and open problem in
valuation (see \cite{Schwartz} (2011)). Motivated by the problem
of coherent risk measures under the volatility uncertainty (see
\cite{Artzner2} (1999)), Peng develops the process with volatility
uncertainty, which is called $G$-Brownian motion in sublinear
expectation space $(\Omega,\mathcal{H},\hat{E})$. He constructs
the G-framework which is a very powerful and beautiful tool to
analyse the uncertainty risk (see \cite{PengB}, \cite{PengC}, and
\cite{PengD}).In the sublinear expectation space the G-Brownian
motion is a G-martingale under the G-expectation, the market
modelled by the G-Brownian motion is incomplete. Using G-framwork,
Epstein and Ji \cite{Epstein} study the utility uncertainty
application in economics, and Chen \cite{Chen} gives a time
consistent G-expectation bid-ask dynamic pricing mechanism for the
European contingent claim in the uncertainty financial market
modelled by SDE driven by the G-Brownian motion.

In this paper we consider to develop a fractional G-white noise
theory under uncertainty. We define fractional G-Brownian motion
(fGBm) $B_H(t)$ with Hurst index $H\in (0,,1)$ in a G-white noise
space, which is a centered G-Gausian process (see Peng
\cite{PengG}) with stationary increment in the sense of
sub-linearity, and it is more realistic to model the financial
market by using the fGBm. Meanwhile, we construct wavelet
decomposition of fGBm on the family of wavelet with compactly
support. We develop a fractional G-white noise theory in a
sublinear expectation space (or G-white noise space)
$(S^{\prime}(R), S(R),\hat{E})$, consider fGBm on the G-white
noise space, define fractional G-noise and set up fractional
G-It\^o-Wick stochastic integral with respect to fGBm. We derive
the fractional G-It\^o formula, define the fractional Malliavin
differential derives, and prove the fractional G-Clark-Ocone
formula. Furthermore, we present the G-Girsanov's Theorem.
Applying our theory in the financial market modelled by
G-Wick-It\^o type stochastic differential equations driven by fGBm
$B_H(t)$, we prove that the sublinear expectation of the
discounted European contingent claim is the bid-ask price of the
European claim.

Our paper is organized as follows: In Sec. 2 we define the fGBm
with Hurst index $H\in (0,1)$ in the sublinear space. We prove
that the fGBm is a continuous stochastic path with H\"older
exponent in $[0,H)$, and has self-similarity property and long
rang dependence property in the sense of the sub-linearity.
Furthermore, we establish the wavelet decomposition for the fGBm
by using wavelets with compactly support. In Sec. 3 we present
fractional G-white noise theory. In Sec. 4 we present the
G-Girsanov's Theorem. In Sec. 5, we apply our theory in the
financial market modelled by G-Wick-It\^o type stochastic
differential equations driven by fGBm $B_H(t)$, and derive the
bid-ask price for the European contingent claim.

\section{Fractional G-Brownian Motion}

\subsection{Sublinear Expectation and Fractional G-Brownian Motion}

Let $\Omega$ be a given set and let $\cal{H}$ be a linear space of
real valued functions defined on $\Omega$ containing constants.
The space $\cal{H}$ is also called the space of random variables.

\begin{definition}
A sublinear expectation $\hat{E}$ is a functional
$\hat{E}:\mathcal{H}\longrightarrow R$ satisfying

(i) Monotonicity:
$$
\hat{E}[X]\geq \hat{E}[Y]\ \ \mbox{if}\ \ X\geq Y.
$$

(ii) Constant preserving:
$$
\hat{E}[c]=c\ \ \mbox{for}\ \ c\in R.
$$

(iii) Sub-additivity: For each $X,Y\in \cal{H}$,
$$
\hat{E}[X+Y]\leq \hat{E}[X]+\hat{E}[Y].
$$

(iv) Positive homogeneity:
$$
\hat{E}[\lambda X]=\lambda\hat{E}[X]\ \ \mbox{for}\ \ \lambda\geq
0.
$$
The triple $(\Omega,\mathcal{H},\hat{E})$ is called a sublinear
expectation space.
\end{definition}

In this paper, we mainly consider the following type of sublinear
expectation spaces $(\Omega,\mathcal{H},\hat{E})$: if
$X_1.X_2,\dots,X_n\in\cal{H}$ then $\varphi(X_1.X_2,\dots,X_n)\in
\cal{H}$ for $\varphi\in C_{b,Lip}(R^n)$, where $C_{b,Lip}(R^n)$
denotes the linear space of functions $\phi$ satisfying
\begin{eqnarray*}
 |\phi(x)-\phi(y)|&\leq& C(1+|x|^m+|y|^m)|x-y| \mbox{ for } x,y\in
 R,\\
&& \mbox{ some } C
> 0, m\in N\mbox{ is depending on }\phi.
\end{eqnarray*}

For each fixed $p\geq 1$, we take $\mathcal{H}_0^p=\{X\in
\mathcal{H},\hat{E}[|X|^p]=0\}$ as our null space, and denote
$\mathcal{H}/\mathcal{H}_0^p$ as the quotient space. We set
$\|X\|_p:=(\hat{E}[|X|^p])^{1/p}$, and extend
$\mathcal{H}/\mathcal{H}_0^p$ to its completion
$\widehat{\cal{H}}_p$ under $\|\cdot\|_p$. Under $\|\cdot\|_p$ the
sublinear expectation $\hat{E}$ can be continuously extended to
the Banach space $(\widehat{\mathcal{H}}_p,\|\cdot\|_p)$. Without
loss generality, we denote the Banach space
$(\widehat{\mathcal{H}}_p,\|\cdot\|_p)$ as
$L^p_G(\Omega,\cal{H},\hat{E})$. For the G-framework of sublinear
expectation space, we refer to \cite{Peng3}, \cite{PengB},
\cite{PengA}, \cite{PengC}, \cite{PengD} and \cite{PengG}. In this
paper we assume that $\underline{\mu}, \overline{\mu},
\underline{\sigma}$ and $\overline{\sigma}$ are nonnegative
constants such that $\underline{\mu}\leq \overline{\mu}$ and
$\underline{\sigma}\leq\overline{\sigma}$.
\begin{definition} Let $X_1$ and $X_2$ be two random variables in a
sublinear expectation space $(\Omega,\mathcal{H},\hat{E})$, $X_1$
and $X_2$ are called identically distributed, denoted by
$X_1\stackrel{d}{=}X_2$ if
\begin{eqnarray*}
\hat{E}[\phi(X_1)]=\hat{E}[\phi(X_2)]& \mbox{for  } \forall\phi\in
C_{b,Lip}(R^n).
\end{eqnarray*}
\end{definition}
\begin{definition}
In a sublinear expectation space $(\Omega,\mathcal{H},\hat{E})$, a
random variable $Y$ is said to be independent of another random
variable $X$, if
\begin{eqnarray*}
\hat{E}[\phi(X,Y)]=\hat{E}[\hat{E}[\phi(x,Y)]|_{x=X}].
\end{eqnarray*}
\end{definition}
\begin{definition} (G-normal distribution) A random variable $X$
on a sublinear expectation space $(\Omega,\mathcal{H},\hat{E})$ is
called G-normal distributed if
\begin{eqnarray*}
aX+b\bar{X}=\sqrt{a^2+b^2}X&\mbox{for  } a,b\ge 0,
\end{eqnarray*}
where $\bar{X}$ is an independent copy of $X$.
\end{definition}

\begin{remark}
For a random variable $X$ on the sublinear space
$(\Omega,\mathcal{H},\hat{E})$, there are four typical parameters
to character $X$
\begin{eqnarray*}
\overline{\mu}=\hat{E}X,&\underline{\mu}=-\hat{E}[-X],\\
\overline{\sigma}^2=\hat{E}X^2,&\underline{\sigma}^2=-\hat{E}[-X^2],
\end{eqnarray*}
where $[\underline{\mu},\overline{\mu}]$ and
$[\underline{\sigma}^2,\overline{\sigma}^2]$ describe the
uncertainty of the mean and the variance of $X$, respectively.

It is easy to check that if $X$ is G-normal distributed, then
$$
\overline{\mu}=\hat{E}X=\underline{\mu}=-\hat{E}[-X]=0,
$$
and we denote the G-normal distribution as
$N(\{0\},[\underline{\sigma}^2,\overline{\sigma}^2])$. If $X$ is
maximal distributed, then
$$
\overline{\sigma}^2=\hat{E}X^2=\underline{\sigma}^2=-\hat{E}[-X^2]=0,
$$
and we denote the maximal distribution as
$N([\underline{\mu},\overline{\mu}],\{0\})$.
\end{remark}
\begin{definition}
We call $(X_t)_{t\in R}$ a d-dimensional stochastic process on a
sublinear expectation space $(\Omega,\mathcal{H},\hat{E})$, if for
each $t\in R$, $X_t$ is a d-dimensional random vector in
$\cal{H}$.
\end{definition}

\begin{definition}
Let $(X_t)_{t\in R}$ and $(Y_t)_{t\in R}$ be d-dimensional
stochastic processes defined on a sublinear expectation space
$(\Omega,\mathcal{H},\hat{E})$, for each
$\underline{t}=(t_1,t_2,\dots,t_n)\in \mathcal{T}$,
$$F_{\underline{t}}^X[\varphi]:=\hat{E}[\varphi(X_{\underline{t}})],\ \ \forall \varphi \in C_{l,Lip}(R^{n\times d})
$$
is called the finite dimensional distribution of $X_t$. $X$ and
$Y$ are said to be indentically distributed, i.e., $
X\stackrel{d}{=}Y $, if
$$
F_{\underline{t}}^X[\varphi]=F_{\underline{t}}^Y[\varphi],\ \ \ \
\forall \underline{t}\in \mathcal{T}\ \ \mbox{and}\ \ \forall
\varphi\in C_{l.Lip}(R^{n\times d})
$$
where $\mathcal{T}:=\{\underline{t}=(t_1,t_2,\dots,t_n): \forall
n\in N,t_i\in R,t_i\neq t_j, 0\leq i,j\leq n,i\neq j \}$.
\end{definition}

\begin{definition}\label{Dbm1}
A process $(B_t)_{t\ge 0}$ on the sublinear expectation space
$(\Omega,\mathcal{H},\hat{E})$ is called a G-Brownian motion if
the following properties are satisfied:

(i) $B_0(\omega)=0$;

(ii) For each $t, s>0$, the increment $B_{t+s}-B_{t}$ is G-normal
distributed by
$N(\{0\},[s\underline{\sigma}^2,s\overline{\sigma}^2]$ and is
independent of $(B_{t_1},B_{t_2},\dots,B_{t_n})$, for each $n\in
N$ and $t_1, t_2,\dots,t_n\in (0, t]$;
\end{definition}

\begin{definition}
A process $(X_t)_{t\in R}$ on a sublinear expectation space
$(\Omega,\mathcal{H},\hat{E})$ is called a centered G-Gaussian
process if for each fixed $t\in R$, $X_t$ is G-normal distributed
$N(\{0\},[\underline{\sigma}_t^2,\overline{\sigma}_t^2])$, where
$0\leq \underline{\sigma}_t\leq \overline{\sigma}_t$.
\end{definition}
\begin{remark} Peng in \cite{PengC} constructs G-framework,
which is a powerful and beautiful analysis tool for risk measure
and pricing under uncertainty. In \cite{PengG}, Peng defines
G-Gaussian processes in a nonlinear expectation space, q-Brownian
motion under a complex-valued nonlinear expectation space, and
presents a new type of Feynman-Kac formula as the solution of a
Schr$\ddot{o}$dinger equation.
\end{remark}
From now on, in this section we start to define a two-sided
G-Brownian motion and a fractional G-Brownian motion, furthermore
we construct the fractional G-Brownian motion and present the
similarity property and long rang dependent property for the
fractional G-Brownian motion in the sense of linearity.
\begin{definition}\label{Dbm2}
A process $(B_{\frac{1}{2}}(t))_{t\in R}\in \Omega$ on the
sublinear expectation space $(\Omega,\mathcal{H},\hat{E})$ is
called a two-sided G-Brownian motion if for two independent
G-Brownian motions $(B^{(1)}_t)_{t\geq 0}$ and $(B^{(2)}_t)_{t\geq
0}$
\begin{eqnarray}
B_{\frac{1}{2}}(t)=\left\{\begin{array}{ll} B^{(1)}(t)& t\geq 0\\
B_{(2)}(-t)&t\le 0\end{array}\right.
\end{eqnarray}
\end{definition}

We consider a family of continuous process under uncertainty which
is corresponding with the fractional Brownian motion (fBm)
provided by Kolmogorov (see \cite{Kol40}) and Manbrot (see
\cite{MvN68}), and we define it as fractional G-Brownian motion
(fGBm):
\begin{definition}\label{Dfgbm}
Let $H\in (0,1)$, a centered G-Gaussian process $(B_H(t))_{t\in
R}$ on the sublinear space $(\Omega,\mathcal{H},\hat{E})$ is
called fractional G-Brownian motion with Hurst index H if

(i) $B_H(0)=0$;

(ii)
\begin{eqnarray}\label{cov}
\left\{\begin{array}{rcl}
\hat{E}[B_H(s)B_H(t)]&=&\frac{1}{2}\overline{\sigma}^2(|t|^{2H}+|s|^{2H}-|t-s|^{2H}),\
\ s,t\in R^+,\\
-\hat{E}[-B_H(s)B_H(t)]&=&\frac{1}{2}\underline{\sigma}^2(|t|^{2H}+|s|^{2H}-|t-s|^{2H}),\
\ s,t\in R^+,
\end{array}
\right.
\end{eqnarray}
we denote the fractional G-Brownian motion as fGBm.
\end{definition}
We can easily check that $(B_{\frac{1}{2}}(t))_{t\in R}$ is
G-Brownian motion, and we denote $B(t)=B_{\frac{1}{2}}(t)$.
\subsection{Moving Average Representation}
Similar with the Mandelbrot-Van Ness representation of fBm, we
give the moving average representation of fGBm with respect to the
G-Brownian motion as follow
\begin{theorem}\label{th_represent}
Let $H\in (0,1)$, for $t\in R$ the Fractional G-Brownian Motion
with Hurst index H is represented as
\begin{eqnarray}\label{Efgbm}
B_H(t,\omega)=C_H^w\int_{R}[(t-s)_+^{H-1/2}-(-s)_+^{H-1/2}]dB(s,\omega),
\end{eqnarray}
where $C_H^w=\displaystyle\frac{(2H\sin{\pi H
}\Gamma(2H))^{1/2}}{\Gamma(H+1/2)}$ and $(B_t)_{t\in R}$ is a
two-sided G-Brownian motion.
\end{theorem}
{\bf Proof.} It is clear that $B_H(0)=\hat{E}[B_H(t)]=0$, and it
is trivial to prove the equations in $(\ref{cov})$ for $s=t$.

From the Definition $\ref{Dbm1}$ and $\ref{Dbm2}$, and by using
G-It\^o stochastic integral (\cite{PengC}) and the integral
transform we have that, for $s>t$
\begin{eqnarray*}
&&\hat{E}[B_H(s)B_H(t)]\\
&=&\frac{2\pi\sin\pi
H\Gamma(2H)}{\Gamma^2(H+\frac{1}{2})}\overline{\sigma}^2
\{\int_{-\infty}^0[(s-u)^{H-1/2}-(-u)^{H-1/2}][(t-u)^{H-1/2}-(-u)^{H-1/2}]du\\
&&+\int_{0}^t(s-u)^{H-1/2}(t-u)^{H-1/2}du\}\\
&=&\overline{\sigma}^2(|s|^{2H}+|t|^{2H}-|s-t|^{2H})\\
&& -\frac{2\pi\sin\pi
H\Gamma(2H)}{\Gamma^2(H+\frac{1}{2})}\overline{\sigma}^2
\{\int_{-\infty}^0[(s-u)^{H-1/2}-(-u)^{H-1/2}][(t-u)^{H-1/2}-(-u)^{H-1/2}]du\\
&& +\int_{0}^t(s-u)^{H-1/2}(t-u)^{H-1/2}du\},
\end{eqnarray*}
thus we prove the first equation in $(\ref{cov})$, and other cases
can be proved in a similar way.

We can prove the second equation in $(\ref{cov})$ with
$\hat{E}[\cdot]$ replaced by $-\hat{E}[-\cdot]$ in the above
equation, hence we prove $(\ref{cov})$.$\ \ \ \ \square$

\subsection{Properties of Fractional G-Brownian Motion}
\begin{definition}
In the sublinear expectation space $(\Omega,\mathcal{H},\hat{E})$,
a process $(X_t)_{t\in R}$ is called to have H-self-similarity
property, if
\begin{eqnarray}
X(at)\stackrel{d}{=}a^{H}X(t)&\mbox{for  } a>0.
\end{eqnarray}
\end{definition}
\begin{theorem}  A fGBm $B_H(t)$ with Hurst index $H\in (0,1)$ in
$(\Omega,L_G^p(\Omega),\hat{E})$ has the following properties

(i) H-self-similar property
\begin{eqnarray*}
B_H(at)\stackrel{d}{=}a^{H}B_H(t)&\mbox{for  } a>0.
\end{eqnarray*}

(ii) The fGBm $(B_H(t))_{t\in R}$ is a continuous path with
stationary increment, with $\beta\in [0,H)$ order H$\ddot{o}$lder
continuous and almost nowhere H$\ddot{o}$lder continuous with
order $\gamma>H$, i.e., for $\alpha\ge 0$,
\begin{eqnarray*}
\hat{E}[|B_H(s)-B_H(t)|^{\alpha}]=
\hat{E}[|B_H(1)|^{\alpha}]|t-s|^{\alpha H},\\
-\hat{E}[-|B_H(s)-B_H(t)|^{\alpha}]=
-\hat{E}[-|B_H(1)|^{\alpha}]|t-s|^{\alpha H}.
\end{eqnarray*}
\end{theorem}
{\bf Proof.}

(i) From the Definition $\ref{Dfgbm}$, the fGBm is a centered
G-Gaussian process, and from
$\hat{E}[B^2_H(at)]=a^{2H}\hat{E}[B^2_H(t)]$ and
$-\hat{E}[-B^2_H(at)]=-a^{2H}\hat{E}[-B^2_H(t)]$ we prove the
H-self-similar property.

(ii) It is easy to check that $B_H(s)-B_H(t)$ and $B_H(s-t)$ is
identity distributed with G-normal distribution
$N(\{0\},[\underline{\sigma}^2(t-s)^{2H},\overline{\sigma}^2(t-s)^{2H}])$,
and the fGBm $(B_H(t))_{t>0}$ has the self similarity property,
therefor, we derive that
\begin{eqnarray*}
\hat{E}[|B_H(s)-B_H(t)|^{\alpha}]&=&\hat{E}[|B_H(s-t)|^{\alpha}]\\
&=&\hat{E}[|B_H(1)|^{\alpha}]|s-t|^{\alpha H},
\end{eqnarray*}
with the similar argument for $-\hat{E}[-\cdot]$, we prove the
theorem. $\ \ \square$

\begin{theorem}\label{longmemory} (Long range dependence) For the fGBm $(B_H(t))_{t\in
R}$ with Hurst index $H$ in sublinear expectation space
$(\Omega,\mathcal{H},\hat{E})$

(i) For $H\in (0,1)$
\begin{eqnarray}\label{co-co1}
\hat{E}[(B_H(n+1)-B_H(n))B_H(1)]&=&
\frac{1}{2}\overline{\sigma}^2[(n+1)^{2H}-2n^{2H}+(n-1)^{2H}],
\end{eqnarray}
\begin{eqnarray}\label{co-co2}
-\hat{E}[-(B_H(n+1)-B_H(n))B_H(1)]&=&
\frac{1}{2}\underline{\sigma}^2[(n+1)^{2H}-2n^{2H}+(n-1)^{2H}].
\end{eqnarray}
(ii) If $H\in (1/2,1)$, there exhibits long rang dependence ,
i.e.,
$$0<\underline{r}(n)<\overline{r}(n),\ \ \ \ \forall n\in N,$$
if $H=1/2$ there exhibits uncorrelated , i.e.,
$$\overline{r}(n)=\underline{r}(n)=0,$$
and if $H\in (0,1/2)$
$$
\lim_{n\longrightarrow
\infty}\underline{r}(n)=\lim_{n\longrightarrow
\infty}\overline{r}(n)=0,
$$
where
$$\overline{r}(n)=\hat{E}[(B_H(n+1)-B_H(n))B_H(1)]$$
and
$$\underline{r}(n)=-\hat{E}[-(B_H(n+1)-B_H(n))B_H(1)]$$
are upper and lower auto-correlation function of
$B_H(n+1)-B_H(n)$, respectively.
\end{theorem}
{\bf Proof.} (i) From the construction of fGBm in the next section
(see $(\ref{fGBm})$ in the next section) we have
\begin{eqnarray}\label{co-cpo1}
\hat{E}[(B_H(n+1)-B_H(n))B_H(1)]&=&\overline{\sigma}^2\int_R[M_HI_{[n,n+1]}(x)M_HI_{[0,1]}(x)]dx,
\end{eqnarray}
\begin{eqnarray}\label{co-cpo2}
-\hat{E}[-(B_H(n+1)-B_H(n))B_H(1)]&=&\underline{\sigma}^2\int_R[M_HI_{[n,n+1]}(x)M_HI_{[0,1]}(x)]dx.
\end{eqnarray}

Define
$$
C_H^{\prime}=[\sin{(\pi H)}\Gamma (2H+1)]^{-1/2}C_H,
$$
where
$$
C_H=[2\Gamma(H-\frac{1}{2})\cos{(\frac{1}{2}\pi(H-\frac{1}{2}))}]^{-1}[\sin{(\pi
H )}\Gamma(2H+1)]^{1/2},
$$
similar with the Definition $\ref{operatMH}$ in the next section
for the operator $M_H$, we denote $M_H^{\prime}$ as the operator
with replace $C_H$ by $C_H^{\prime}$ in the definition $M_H$. For
$0\leq a<b$ and $0<H<1$, by Parseval's Theorem
\begin{eqnarray*}
\int_R[M^{\prime}_HI_{[a,b]}(x)]^2dx&=&\displaystyle\frac{1}{2\pi}\int_R[\widehat{M^{\prime}_HI_{[a,b]}}(\xi)]^2d\xi\\
&=&\displaystyle\frac{1}{2\pi}\int_R|\xi|^{1-2H}\widehat{I_{[a,b]}}(\xi)^2d\xi\\
&=&\displaystyle\frac{1}{2\pi}\int_R|\xi|^{1-2H}[\displaystyle\frac{e^{-ib\xi}-e^{-ia\xi}}{-i\xi}]^2d\xi\\
&=&\displaystyle\frac{1}{\sin{\pi H}\Gamma(2H+1)}(b-a)^{2H},
\end{eqnarray*}
from which and notice $(\ref{co-cpo1})$ and $(\ref{co-cpo2})$ we
can prove (i).

(ii) From $(\ref{co-co1})$ we have that
\begin{eqnarray*}
\overline{r}_n\sim H(2H-1)n^{2H-2}\hat{E}[B_H^2(1)],\ \
n\longrightarrow \infty,\ \ H\neq 1/2,\\
\overline{r}_n=0,\ \ H=1/2.
\end{eqnarray*}
and
\begin{eqnarray*}
\sum_{n=1}^{\infty}\overline{r}_n=\hat{E}[B_H^2(1)]\lim_{n\longrightarrow
\infty }((n+1)^{2H}-n^{2H}-1)\left\{\begin{array}{ll}<\infty,&H\in
(0.1/2);\\
=0;&H=1/2;\\
=\infty,&H\in(1/2,1),\end{array}\right.
\end{eqnarray*}
we can also derive the similar expressions for $\underline{r}_n$,
thus we finish the proof.$\ \ \square$

\subsection{Wavelet Decomposition of Fractional G-Brownian Motion}
We consider to expand the fGBm on the periodic compactly supported
wavelet family (see \cite{Daubechies} and \cite{Hemandez}):
\begin{eqnarray}
\{\psi_{j,k}:x\longrightarrow \sum_{l\in Z}\psi(2^j(x-l)-k),\ \
j\geq 0,\ \ 0\leq k\leq 2^j-1\}
\end{eqnarray}
where $\psi$ is a mother wavelet, and we denote $\phi(x)$ as its
periodic scaling function. We assume that

\begin{itemize}
\item the wavelet $\psi$ belongs to the Schwartz class $S(R)$;
\item the $\psi$ has $N(\geq 2)$ vanishing moments, i.e.,
$$\int_{-\infty}^{\infty}t^N\psi(t)dt=0.$$
\end{itemize}

By convention, if $j=-1$, $0\leq k\leq 2^{-1}-1$ means $k=0$, we
denote
$$2^{-\frac{1}{2}}\psi_{-1,k}(t)=\phi(x-k), 0\leq k\leq 2^{-1}-1.$$

Then the periodized wavelet family
\begin{eqnarray}\label{basiswavelet}
\{ 2^{j/2}\psi_{j,k}(t),\ j\geq -1,\ 0\leq k\leq 2^j-1\}
\end{eqnarray}
form an orthonormal basis of $L^2(T)$, where $T:=R/Z$ (1-period).
Without loss of generality, we consider $T=[0,1]$.

For $\alpha>0$, we denote Liouville fractional integral as
\begin{eqnarray}\label{RLI}
(I^{\alpha}f)(x):=\displaystyle\frac{1}{\Gamma(\alpha)}\int_0^t(t-x)^{\alpha-1}f(x)dx,
\end{eqnarray}
and define Riemann-Liouville fractional integral coincide with the
Marchaud fractional integral as follows
\begin{eqnarray}\label{IM}
(I_M^{\alpha}f)(x):=\displaystyle\frac{1}{\Gamma(\alpha)}\int_{-\infty}^{+\infty}[(t-x)_+^{\alpha-1}-(-x)_+^{\alpha-1}]f(x)dx,
\end{eqnarray}

\begin{theorem}
There exists a wavelet expansion for a fGBm process $B_H(t)$,
i.e., for $H\in (0,1)$
\begin{eqnarray}\label{expandfgbm}
B_H(t)=C_H^w\sum_{j=-1}^{\infty}\sum_{k=0}^{2^{j}-1}\mu_{j,k}(I_M^{\alpha}\psi_{j,k})(t)
\end{eqnarray}
where
\begin{eqnarray*}
\alpha&=&H+\frac{1}{2},\\
C_H^w&=&\frac{(2H\sin{\pi H
}\Gamma(2H))^{1/2}}{\Gamma(H+1/2)},\\
\mu_{j,k}&=&2^{-(H-\frac{1}{2})j}\varepsilon_{j,k}
\end{eqnarray*}
and $\varepsilon_{j,k}$ are i.i.d. G-normal distributed with
$B_H(1)\sim N(\{0\},[\underline{\sigma}^2,\overline{\sigma}^2])$.
\end{theorem}
{\bf Proof.} (i) We denote the right hand side of
$(\ref{expandfgbm})$ as
$$F(t)=C_H^w\sum_{j=-1}^{\infty}\sum_{k=0}^{2^{j}-1}\mu_{j,k}(I_M^{\alpha}\psi_{j,k})(t)$$.
Without loss generality, we can rewrite $F(t)$ as follows
\begin{eqnarray}
F(t)=C_H^w\sum_{n=-1}^{\infty}f_n(t)\varepsilon_n,
\end{eqnarray}
where  $\{f_n\}_{n=-1}^{\infty}$ denotes the countable Riesz basis
$\{2^{-(H+1)j}(I_M^{\alpha}\psi_{j,k})(x)\}_{j\geq
-1,k=0,1,\cdots,2^{j}-1}$ of $L^2(R)$ (see \cite{Unser}).

For proving that the right-hand side of above equation
 defines a generalized process, i.e., as a linear functional
 \begin{eqnarray}
 F(u)=\int_{-\infty}^{\infty}F(t)\overline{u(t)}dt, \ \ \mbox{for }
 \forall u\in S(R),
 \end{eqnarray}
 we only need to prove
 \begin{eqnarray}
 \|F\|_{H^{-1}}=\|C_H^w\sum_{n=-1}^{\infty}f_n(t)\varepsilon_n\|_{H^{-1}}<
 \infty.
 \end{eqnarray}
By the representation theorem of a sublinear expectation (see
\cite{PengC}), there exists a family of linear expectations
$\{E_{\theta}:\theta\in \Theta\}$ such that
\begin{eqnarray}
\hat{E}[X]=\sup_{\theta\in \Theta}E_{\theta}[X], \ \ \mbox{for}\
X\in \mathcal{H}.
\end{eqnarray}
Thus, by Kolmogrov's convergence critera we conclude that
\begin{eqnarray}
\|C_H^w\sum_{n=-1}^{\infty}f_n(t)\varepsilon_n\|_{H^{-1}}<
 \infty.
\end{eqnarray}
Consequently, by Plancherel theorem
\begin{eqnarray*}
|<F,u>|&=&\frac{1}{2\pi}|\int_{-\infty}^{\infty}\widehat{F}(\xi)\overline{\widehat{u}(\xi)}d\xi|\\
&=&\frac{1}{2\pi}|\int_{-\infty}^{\infty}\hat{F}(\xi)(1+\xi^2)^{-1/2}\overline{\widehat{u}(\xi)}(1+\xi^2)^{1/2}d\xi|\\
&\leq & \frac{1}{2\pi}\|F\|_{H^{-1}}\|u\|_{H^{-1}}\\
&< &\infty.
\end{eqnarray*}
where $\hat{u}(\xi)=\int_{-\infty}^{\infty}u(t)e^{-it\xi}dt$ is
Fourier transform of $u$.

(ii) We prove that $\{B_H(t),t\in R\}$ is a centered generalized
G-Gaussian process with stationary increment, i.e., with zero mean
and
\begin{eqnarray}\label{increment}\begin{array}l
\hat{E}B_H(u)\overline{B_H}(v)=\displaystyle\frac{\overline{\sigma}}{2}\int_{-\infty}^{\infty}\int_{-\infty}^{\infty}(|t|^{2H}+|s|^{2H}-|t-s|^{2H})u(t)\overline{v(s)}dtds,\\
-\hat{E}[-B_H(u)\overline{B_H}(v)]=\displaystyle\frac{\underline{\sigma}}{2}\int_{-\infty}^{\infty}\int_{-\infty}^{\infty}(|t|^{2H}+|s|^{2H}-|t-s|^{2H})u(t)\overline{v(s)}dtds.
\end{array}
\end{eqnarray}

From the definition of the fractional integral $(\ref{IM})$, we
have
\begin{eqnarray*}
B_H(u)&=&C_H^w\int_{-\infty}^{\infty}\sum_{j=-1}^{\infty}\sum_{k=0}^{2^j-1}2^{-(H+\frac{1}{2})j}\varepsilon_{j,k}(I_M^{\alpha}\psi_{j,k}(t)\overline{u}(t)dt\\
&=&C_H^w\sum_{j=-1}^{\infty}\sum_{k=0}^{2^j-1}2^{-(H+\frac{1}{2})j}\varepsilon_{j,k}\int_{-\infty}^{\infty}\overline{u}(t)\int_{-\infty}^{\infty}\left[((I^{\alpha}\delta)(t-x))_+-((I^{\alpha}\delta)(-x))_+\right]\psi_{j,k}(x)dxdt,
\end{eqnarray*}
where
$(I^{\alpha}\delta)(t-s)=\displaystyle\frac{(t-s)^{\alpha-1}}{\Gamma(\alpha)}$.

G-normal distributed $\varepsilon_{j,k}\  (j=-1,0,1,\dots,;
k=0,\dots,2^j-1)$ are independent, we derive
\begin{eqnarray*}
&&\hat{E}B_H(u)\overline{B_H}(v)\\
&=&\hat{E}[B_H^2(1)](C_H^w)^2\sum_{j=-1}^{\infty}\sum_{k=0}^{2^j-1}2^{-(2H+1)j}\int_R\int_R\left[((I^{\alpha}\delta)(t-s))_+-((I^{\alpha}\delta)(-s))_+\right]u(t)\psi_{j,k}(s)dsdt\\
&\ &\ \ \ \ \ \ \ \ \ \ \ \ \ \ \ \ \ \ \ \ \ \ \ \ \ \ \ \ \ \
\int_R\int_R\left[((I^{\alpha}\delta)(t-s))_+-((I^{\alpha}\delta)(-s))_+\right]\overline{v}(t)\psi_{j,k}(s)dsdt\\
&=&\overline{\sigma}^2(C_H^w)^2\int_R\Big[\int_R\left[((I^{\alpha}\delta)(t-s))_+-((I^{\alpha}\delta)(-s))_+\right]u(t)dt\\
&\ &\ \
\int_R\overline{v}(t)\sum_{j=-1}^{\infty}\sum_{k=0}^{2^j-1}<\left[((I^{\alpha}\delta)(t-s^{\prime}))_+-((I^{\alpha}\delta)(-s^{\prime}))_+\right],2^{-(H+\frac{1}{2})j}\psi_{j,k}(s^{\prime})>_{L^2(R)}2^{-(H+\frac{1}{2})j}\psi_{j,k}(s)dt\Big]ds\\
&=&\overline{\sigma}^2(C_H^w)^2\int_R\Big[\int_R\left[((I^{\alpha}\delta)(t-s))_+-((I^{\alpha}\delta)(-s))_+\right]u(t)dt
\int_R\overline{v}(t)\left[((I^{\alpha}\delta)(t-s))_+-((I^{\alpha}\delta)(-s))_+\right]dt\Big]ds\\
&=&\overline{\sigma}^2(C_H^w)^2\int_R\int_Ru(t)\overline{v}(s)
\Big[\int_R\left[((I^{\alpha}\delta)(t-t^{\prime}))_+-((I^{\alpha}\delta)(-t^{\prime}))_+\right]
\left[((I^{\alpha}\delta)(s-t^{\prime}))_+-((I^{\alpha}\delta)(-t^{\prime}))_+\right]dt^{\prime}\Big]dsdt.
\end{eqnarray*}
Following from the proof of Theorem $\ref{th_represent}$, we have
\begin{eqnarray*}
&&\overline{\sigma}^2(C_H^w)^2\int_R\left[((I^{\alpha}\delta)(t-t^{\prime}))_+-((I^{\alpha}\delta)(-t^{\prime}))_+\right]
\left[((I^{\alpha}\delta)(s-t^{\prime}))_+-((I^{\alpha}\delta)(-t^{\prime}))_+\right]dt^{\prime}\\
&=&\displaystyle\frac{\overline{\sigma}^2}{2}[|t|^{2H}+|s|^{2H}-|t-s|^{2H}],
\end{eqnarray*}
we finish the first equation in $(\ref{increment})$. With the
similar argument, we can prove the second part in
$(\ref{increment})$. Thus, we prove that the right hand of
($\ref{expandfgbm}$) is a generalized fGBm, we finish the proof of
the theorem. $\ \ \square$

\begin{remark}
For construct the G-normal distributed random vector, for example
$B_H(1)$, Peng in \cite{PengC} proposed the central limit theorem
with zero-mean.

Let $\{X_i\}_{i=1}^{\infty}$ be a sequence of $R^d-$ valued random
variables on a sublinear expectation space
$(\Omega,\mathcal{H},\hat{E})$, $\hat{E}[X_1]=-\hat{E}[-X_1]=0$,
and assume that $X_{i+1}\stackrel{d}{=}X_i$ and $X_{i+1}$ is
independence from $\{X_1,\dots,X_{i}\}$. Then
$$
S_n\stackrel{\tt{law}}{\longrightarrow} X,
$$
where
$$
S_n:=\displaystyle\frac{1}{\sqrt{n}}\sum_{i=1}^nX_i
$$
and $X$ is G-normal distributed.
\end{remark}

\section{Fractional G-Noise and Fractional G-It\^o Formula}
\subsection{Fractional G-Brownian Motion on the G-White Noise Space}
Let $S(R)$ denotes the Schwartz space of rapidly decreasing
infinitely differentiable real valued functions, let
$S^{\prime}(R)$ be the dual space of $S(R)$, and $<\cdot,\cdot>$
denotes the dual operation, for $f\in L^2(R)$ by approximating by
step functions
\begin{eqnarray}
<f,\omega>:=\int fdB(\omega),
\end{eqnarray}
where $B(\omega)=B(\cdot,\omega)$ is the two-sided G-Brownian
motion with $B(1)\sim
N(\{0\},[\underline{\sigma}^2,\overline{\sigma}^2])$. Then
$(S^{\prime}(R),S(R),\hat{E})$ is a sublinear expectation space.

\begin{remark}
Concerning the G-framework and G-It\^o stochastic integral
theorem, we refer to Peng's paper \cite{PengA}, book \cite{PengC}
and references therein.
\end{remark}

Denote $I_{[0,t]}(s)$ as the indicator function
\begin{eqnarray}
I_{[0,t]}(s)=\left\{\begin{array}{ll} 1& \mbox{if  } 0\leq s\leq
t\\
-1& \mbox{if  } t\leq s\leq
0\\
0& \mbox{otherwise  }
\end{array}
\right.
\end{eqnarray}
Define the following process
\begin{eqnarray}\label{FGBM2}
\tilde{B}_t(\omega):=<I_{[0,t]}(\cdot),\omega>,
\end{eqnarray}
then $(\tilde{B}_t)_{t\in R}$ is two-sided G-Brownian motion with
$\tilde{B}_t\sim
N(\{0\},[\underline{\sigma}^2|t|,\overline{\sigma}^2|t|])$.
Without loss generality, for $t\in R$ we denote $B_t$ as two-sided
G-Brownian motion $\tilde{B}_t$.

For $H\in (0,1)$, we define the following operator $M_H$
\begin{definition}\label{operatMH}
The operator $M_{H}$ is defined on functions $f\in S(R)$ by
\begin{eqnarray}
\widehat{M_Hf}(y)=|y|^{1/2-H}\hat{f}(y),&y\in R,
\end{eqnarray}
where
$$
\hat{g}:=\int_Re^{-ixy}g(x)dx
$$
denotes the Fourier transform.
\end{definition}
For $0<H<\frac{1}{2}$ we have
\begin{eqnarray}
M_Hf(x)=C_H\int_R\displaystyle\frac{f(x-t)-f(x)}{|t|^{3/2-H}}dt,
\end{eqnarray}
where
$$
C_H=[2\Gamma(H-\frac{1}{2})\cos{(\frac{1}{2}\pi(H-\frac{1}{2}))}]^{-1}[\sin{(\pi
H )}\Gamma(2H+1)]^{1/2}.
$$

For $H=\frac{1}{2}$ we have
\begin{eqnarray}
M_Hf(x)=f(x).
\end{eqnarray}

For $\frac{1}{2}<H<1$ we have
\begin{eqnarray}
M_Hf(x)=C_H\int_R\displaystyle\frac{f(t)}{|t-x|^{3/2-H}}dt.
\end{eqnarray}

We define
\begin{eqnarray}
L_H^2(R)&:=&\{f: M_Hf\in L^2(R)\}\nonumber\\
 &=&\{f: |y|^{\frac{1}{2}-H}\hat{f}(y)\in L^2(R)\}\\
 &=&\{f:\|f\|_{L^2_H(R)}<\infty\},\ \ \mbox{where }
 \|f\|_{L_H^2(R)}=\|M_Hf\|_{L^2(R)},\nonumber
\end{eqnarray}
then the operator $M_H$ can be extended from $S(R)$ to $L_H^2(R)$.

For $H\in (0,1)$, consider the following process
\begin{eqnarray}\label{fGBm}
\tilde{B}_H(t,\omega):=<M_HI_{(0,t)}(\cdot),\omega>
\end{eqnarray}
then, for $t\in R$ it is a centered G-Gaussian process (see Peng
(2011) \cite{PengG}) with
$\tilde{B}_H(0)=\hat{E}[\tilde{B}_H(t)]=0$, and
\begin{eqnarray*}
\hat{E}[\tilde{B}_H(s)\tilde{B}_H(t)]&=&\overline{\sigma}^2[\int_RM_HI_{(0,s)}(x)M_HI_{(0,t)}(x)dx]\\
&=&\frac{1}{2}\overline{\sigma}^2[|t|^{2H}+|s|^{2H}-|s-t|^{2H}],\\
-\hat{E}[-\tilde{B}_H(s)\tilde{B}_H(t)]&=&\underline{\sigma}^2[\int_RM_HI_{(0,s)}(x)M_HI_{(0,t)}(x)dx]\\
&=&\frac{1}{2}\underline{\sigma}^2[|t|^{2H}+|s|^{2H}-|s-t|^{2H}],
\end{eqnarray*}
then the continuous process $\tilde{B}_H(t)$ is a fGBm with Hurst
index $H$ , we denote $\tilde{B}_H(t)$ as $B_H(t)$ .

Let $f(x)=\sum_ja_jI_{[t_j,t_{j+1}]}(x)$ be a step function, then
\begin{eqnarray}
<M_Hf,\omega>=\int_Rf(t)dB_H(t),
\end{eqnarray}
and can be extended to all $f\in L_H^2(R)$. And we also have
\begin{eqnarray}
\int_Rf(t)dB_H(t)=\int_RM_Hf(t)dB(t),\ \ f\in L^2_H(R).
\end{eqnarray}

\subsection{Fractional G-Noise}

Recall the Hermite polynomials
\begin{eqnarray*}
h_n(x)=(-1)^ne^{\frac{x^2}{2}}\displaystyle\frac{d^n}{dx^n}e^{-\frac{x^2}{2}},&n=0,1,2,\cdots\cdot
\end{eqnarray*}
We denote the Hermite functions as follows:
\begin{eqnarray*}
\tilde{h}_n(x)=\pi^{-\frac{1}{4}}((n-1)!)^{-\frac{1}{2}}h_{n-1}(\sqrt{2}x)e^{-\frac{x^2}{2}},\
\ n=1,2,\cdots\cdot
\end{eqnarray*}
Then $\{\tilde{h}_n,n=1,2,\cdots\}$ is an orthonormal basis of
$L^2(R)$ and
\begin{eqnarray*}
|\tilde{h}_n(x)|\leq \left\{\begin{array}{ll}
Cn^{-\frac{1}{12}}&\mbox{if  } |x|\leq 2\sqrt{n}\\
Ce^{-\gamma x^{2}}&\mbox{if  } |x|> 2\sqrt{n},
 \end{array}\right.
\end{eqnarray*}
where $C$ and $\gamma$ are constants independent of $n$. Define
\begin{eqnarray*}
e_i(x):=M_H^{-1}\tilde{h}_i(x),&i=1,2,\cdots\cdot
\end{eqnarray*}
Then $\{e_i,i=1,2,\cdots\}$ is an orthonomal basis of $L_H^2(R)$.

We denote $J$ as the set of all finite multi-indices
$\alpha=(\alpha_1,\cdots,\alpha_n)$ for some $n\ge 1$ and
$\alpha_i\in N_0=\{0,1,2,\cdots\}$, and for $\alpha\in J$
\begin{eqnarray*}
H_{\alpha}(\omega)&=&\Pi_{i=1}^{n}h_{\alpha_i}(<\tilde{h}_i,\omega>)\\
&=&\Pi_{i=1}^{n}h_{\alpha_i}(\int_R\tilde{h}_idB(\omega)),
\end{eqnarray*}
where $(B(t,\omega))_{t\in R}$ is the two-sided G-Brownian motion.

Denote $\tilde{h}^{\hat{\otimes}\alpha}$ as the symmetric product
with factors $\tilde{h}_1,\dots,\tilde{h}_n$ with each
$\tilde{h}_i$ being taken $\alpha_i$ times, similar with the
statement for the fundamental result of It\^o \cite{Ito} (1951) we
denote
\begin{eqnarray}
\int_{R^{|\alpha|}}\tilde{h}^{\hat{\otimes}\alpha}dB^{\otimes|\alpha|}:=H_{\alpha}(\omega).
\end{eqnarray}
\begin{definition} We define space $\hat{L}_H^2(R^n)$ as follows
\begin{eqnarray*}
\hat{L}_H^2(R^n):=\{f(x_1,\dots,x_n)\mbox{ be symmetric function
of }(x_1,\dots,x_n)\left|\right.\ M_H^nf\in L^2(R^n) \},
\end{eqnarray*}
where $M_H^nf$ means the operator $M_H$ is applied to each
variable of $f$, and we denote
\begin{eqnarray}
\|f\|_{\hat{L}_H^2(R^n)}^2:=\int_{R^{n}}(M_H^n f)^2ds.
\end{eqnarray}
For $f\in \hat{L}_{H}^2(R^n)$, we define
$$
\int_{R^n}fdB_{H}^{\otimes n}:=\int_{R^n}(M_H^n f)dB^{\otimes n}.
$$
\end{definition}
\begin{definition}
The random variable
$$F\in
L^2_{G,H}(S^{\prime}(R),S(R),\hat{E}),\ \mbox{if and only if }
F\circ M_H\in L^2_G(S^{\prime}(R),S(R),\hat{E}).
$$
\end{definition}
The expansion of $F\circ M_H$ in terms of the Hermite functions:
\begin{eqnarray*}
F(M_H\omega)&=&\sum_{\alpha}c_{\alpha}H_{\alpha}(\omega)\\
&=&\sum_{\alpha}c_{\alpha}h_{\alpha_1}(<M_He_1,\omega>)\dots
h_{\alpha_n}(<M_He_n,\omega>)\\
&=&\sum_{\alpha}c_{\alpha}h_{\alpha_1}(<e_1,M_H\omega>)\dots
h_{\alpha_n}(<e_n,M_H\omega>).
\end{eqnarray*}
Consequently,
\begin{eqnarray*}
F(\omega)&=&\sum_{\alpha}c_{\alpha}h_{\alpha_1}(<e_1,\omega>)\dots
h_{\alpha_n}(<e_n,\omega>)\\
&=&\sum_{\alpha}c_{\alpha}H_{\alpha}(\omega).
\end{eqnarray*}
For giving $F\in L^2_G(S^{\prime}(R),S(R),\hat{E})$, we have
\begin{eqnarray*}
F(\omega)&=&\sum_{\alpha}c_{\alpha}H_{\alpha}(\omega)\\
&=&\sum_n\sum_{|\alpha|=n}c_{\alpha}\int_{R^n}\tilde{h}^{\hat{\otimes}\alpha}dB^{\otimes
n}\\
&=&\sum_{n}\sum_{|\alpha|=n}c_{\alpha}\int_{R^n}e^{\hat{\otimes}\alpha}dB_{M_H}^{\otimes
n}
\end{eqnarray*}
Aase et al. \cite{Aase} (2000), Holden et al. \cite{Holden}, and
Elliott $\&$ Hoek \cite{Elliott} (2003) gave the definitions of
Hida space $(S)$ and $(S)^*$ under the probability framework. Here
we define G-Hida space $(S)$ and $(S)^*$ as follows
\begin{definition}
(i) We define the G-Hida space $(S)$ to be all functions $\psi$
with the following expansion
\begin{eqnarray}
\Psi(\omega)=\sum_{\alpha\in J}a_{\alpha}\cal{H}_{\alpha}(\omega)
\end{eqnarray}
satisfies
$$\|\Psi\|_{k,(S)}:=\sum_{\alpha\in J}a^2_{\alpha}\alpha !(2N)^{k\alpha}<\infty \mbox{ for
all }k=1,2,\cdots,$$ where
$$
(2N)^{\gamma}:=(2\cdot 1)^{\gamma_1}(2\cdot
2)^{\gamma_2}\cdots(2\cdot m)^{\gamma_m} \mbox{  if  }
\gamma=(\gamma_1,\gamma_2,\cdots,\gamma_m)\in J.
$$
(ii) We define the G-Hida space $(S)^*$ to be the set of the
folowing expansions
\begin{eqnarray}
\Phi(\omega)=\sum_{\alpha\in J}b_{\alpha}\cal{H}_{\alpha}(\omega)
\end{eqnarray}
such that $$\|\Phi\|_{-q,(S)^*}:=\sum_{\alpha\in
J}b_{\alpha}^2\alpha ! (2N)^{-q\alpha}<\infty \mbox{ for some
integer }q\in (0,+\infty).$$
\end{definition}
The duality between $(S)$ and $(S)^*$ is given as follows: for
\begin{eqnarray*}
\Psi(\omega)=\sum_{\alpha\in
J}a_{\alpha}\cal{H}_{\alpha}(\omega)\in (S),\\
\Phi(\omega)=\sum_{\alpha\in
J}b_{\alpha}\cal{H}_{\alpha}(\omega)\in (S)^*
\end{eqnarray*}
\begin{eqnarray*}
<<\Psi,\Phi>>:=\sum_{\alpha\in J}\alpha !a_{\alpha}b_{\alpha}.
\end{eqnarray*}

The space $(S)^*$ is convenient for the wick product
\begin{definition}
If for $F_i(\cdot)\in (S)^*$
$$F_i(\omega)=\sum_{\alpha\in
J}c_{\alpha}^{(i)}H_{\alpha}(\omega),\ \ \mbox{i=}1,2,$$ we define
their Wick product $(F_1\diamond F_2)(\omega)$ by
\begin{eqnarray}
(F_1\diamond F_2)(\omega)=\sum_{\alpha,\beta\in
J}c_{\alpha}^{(1)}c_{\beta}^{(2)}H_{\alpha+\beta}(\omega)=\sum_{\gamma\in
J
}\left(\sum_{\alpha+\beta=\gamma}c_{\alpha}^{(1)}c_{\beta}^{(2)}\right)H_{\gamma}(\omega).
\end{eqnarray}
\end{definition}

From the chaos expansion of the fGBm $B_H(t)$, we have
\begin{eqnarray*}
B_H(t)&=&<M_HI_{[0,t]},\omega>\\
&=&<I_{[0,t]},M_H\omega>\\
&=&<\sum_{k=1}^{\infty}(I_{[0,t]},e_k)_{L_H^2(R)}e_k,M_H\omega>\\
&=&<\sum_{k=1}^{\infty}(M_HI_{[0,t]},M_He_k)_{L^2(R)}e_k,M_H\omega>\\
&=&\sum_{k=1}^{\infty}(M_HI_{[0,t]},\tilde{h}_k)_{L^2(R)}<e_k,M_H\omega>\\
&=&\sum_{k=1}^{\infty}(I_{[0,t]},M_H\tilde{h}_k)_{L^2(R)}<e_k,M_H\omega>\\
&=&\sum_{k=1}^{\infty}\int_0^tM_H\tilde{h}_k(s)ds
H_{\varepsilon^{(k)}}(\omega),
\end{eqnarray*}
where $\varepsilon^{(k)}=(0,0,\cdots,0,1,0,\cdots,0)$ with $1$ on
the $k$th entry, $0$ otherwise, and $k=1,2,\cdots$. We define the
$H$-fractional G-noise $W_H(t)$ by the expansion as follows
\begin{definition} $H$-fractional G-noise with respect to the fGBm $B_H(t)$ is define as following
\begin{eqnarray}
W_H(t)=\sum_{k=1}^{\infty}M_H\tilde{h}_k(t)H_{\varepsilon^{(k)}}(\omega).
\end{eqnarray}

For $H=\frac{1}{2}$, we call $\frac{1}{2}$-fractional G-noise
$W_{\frac{1}{2}}(t)$ with respect to the G-Brownian motion $B(t)$
as G-white noise.

For $\frac{1}{2}<H<1$, $H$-fractional G-noise $W_H(t)$ is called
fractional G-black noise.

For $0<H<\frac{1}{2}$, $H$-fractional G-noise $W_H(t)$ is called
fractional G-pink noise.
\end{definition}
Then it can be shown that $W_H(t)$ for all $t$ as
\begin{eqnarray}
\displaystyle\frac{dB_H(t)}{dt}=W_H(t).
\end{eqnarray}

\begin{remark}
Long Range Dependence Theorem $\ref{longmemory}$ characterizes the
$H$-fractional G-noise, especially, its uncertainty. The
fractional G-black noise has persistence long memory, the
fractional G-pink noise has negative correlations in the sense of
the sub-linearity and quickly alternatively change its value, and
the fractional G-white noise is i.i.d. G-Gaussian sequence. The
characters of fractional G-noise make it has strong natural
background (\cite{Kol41}, \cite{Mandelbrot}, \cite{Peters91} and
\cite{Gha96}).
\end{remark}

\begin{lemma}
The $H$-fractional G-noise $W_H(t)$ is in the G-Hida space $(S)^*$
for all $t$, for $H\in (0,1)$.
\end{lemma}
{\bf Proof.} By the definition $\ref{operatMH}$, and notice that
$\hat{\tilde{h}}_k(y)=\sqrt{2\pi}(-i)^{k-1}\tilde{h}_k(y)$ we
derive that
\begin{eqnarray*}
M_h\tilde{h}_k(t)&=&\frac{1}{2\pi}\int_R
e^{iyt}|y|^{\frac{1}{2}-H}\hat{\tilde{h}}_k(y)dy\\
&=&\frac{(-i)^{k-1}}{\sqrt{2\pi}}\int_R
e^{iyt}|y|^{\frac{1}{2}-H}\tilde{h}_k(y)dy.
\end{eqnarray*}
From Hille \cite{Hille} (1958) and Thangavelu \cite{Thangavelu}
(1993),
$$
|\tilde{h}_k(y)|\leq \left\{\begin{array}{ll} C
k^{-\frac{1}{12}}& \mbox{if } |y|\leq 2\sqrt{k}\\
C e^{-\gamma y^2}& \mbox{if } |y|\geq 2\sqrt{k}
\end{array}
\right.
$$
where $C$ and $\gamma$ are constants independent of $k$ and $y$.
Thus
\begin{eqnarray*}
|M_h\tilde{h}_k(t)|&\leq & C
\left[\int_{-2\sqrt{k}}^{2\sqrt{k}}|y|^{\frac{1}{2}-H}dy\cdot
k^{-\frac{1}{12}}+[(\int_{-\infty}^{-2\sqrt{k}}+\int^{+\infty}_{2\sqrt{k}})|y|^{\frac{1}{2}-H}e^{-\gamma
y^2}dy]\right]\\
&\leq & C k^{-\frac{1}{12}+\frac{3}{4}-\frac{H}{2}}.
\end{eqnarray*}
Therefore, we have
\begin{eqnarray*}
\|W_H(t)\|_{-q,(S)^*}^2&=&\sum_{k=1}^{\infty}|M_H\tilde{h}_k(t)|^2(2k)^{-q}\\
&\leq &C\sum_{k=1}^{\infty}k^{-\frac{1}{6}+\frac{3}{2}-H-q}\\
&\leq &C,\ \ \mbox{bounded uniformly for all } t, \mbox{ and }
q>\frac{4}{3}.
\end{eqnarray*}
Hence, we finish the proof of the Lemma. $\ \ \square$

\begin{definition}
Suppose $Y\in (S)^*$ is such that $Y(t)\diamond W_H(t)$ is
integrable in $(S)^*$, we say that $Y$ is $dB_H$-integrable and we
define the fractional G-It\^o-Wick integral of $Y(t)=Y(t,\omega)$
with respect to $B_H(t)$ by
\begin{eqnarray}
\int_RY(t)dB_H(t):=\int_RY(t)\diamond W_H(t)dt.
\end{eqnarray}
\end{definition}

Using Wick calculus we derive
\begin{lemma}
\begin{eqnarray}
\int_0^T B_H(t)dB_H(t)=\frac{1}{2}(B_H(T))^2-\frac{1}{2}T^{2H}.
\end{eqnarray}
\end{lemma}

\subsection{Fractional G-It\^o Formula}
Consider the fractional stochastic differential equation driven by
fGBm $B_H(t)$
\begin{eqnarray}\label{efsde}
\left\{\begin{array}{rcl}
dX(t)&=&X(t)\diamond(\alpha(t)dt+\beta(t)dB_H(t)),\ \ t\geq 0,\\
X(0)&=&x.
\end{array}
\right.
\end{eqnarray}
We write the above fractional stochastic differential equation in
$(S)^*$
\begin{eqnarray}
\frac{dX(t)}{dt}=X(t)\diamond [\alpha(t)+\beta(t)W_H(t)]dt,
\end{eqnarray}
by Wick product formula we have
\begin{eqnarray}
X(t)&=&X(0)\diamond\exp^{\diamond}\left(\int_0^t\alpha(s)ds+\int_0^t\beta(s)dB_H(s)\right),
\end{eqnarray}
i.e.
\begin{eqnarray}
X(t)&=&x\exp\left(\int_0^t\beta(s)dB_H(s)+\int_0^t\alpha(s)ds-\displaystyle\frac{1}{2}\int_R(M_H(\beta(s)I_{[0,t](s)}))^2ds\right).
\end{eqnarray}
is the solution of the fractional SDE $(\ref{efsde})$.
\begin{theorem}
Assume that $f(s,x)\in C^{1,2}(R\times R)$, the fractional G-It\^o
formula is as follows:
\begin{eqnarray}
f(t,B_H(t))=f(0,0)+\int_0^t\displaystyle\frac{\partial f}{\partial
s}(s,B_H(s))ds+\int_0^t\displaystyle\frac{\partial f}{\partial
x}(s,B_H(s))dB_H(s)+H\int_0^t\displaystyle\frac{\partial^2
f}{\partial x^2}(s,B_H(s))s^{2H-1}ds.
\end{eqnarray}
\end{theorem}
{\bf Proof.} Let
\begin{eqnarray}\label{Gitog}
g(t,x)=\exp(\alpha x+\beta (t))
\end{eqnarray}
where $\alpha\in R$ be a constant and $\beta:R\longrightarrow R$
be a deterministic differentiable function. Set
\begin{eqnarray*}
Y(t)=g(t,B_H(t)).
\end{eqnarray*}
Then
\begin{eqnarray*}
Y(t)=\exp^{\diamond}(\beta(t)+\alpha
B_H(t)+\frac{1}{2}\alpha^2t^{2H}),
\end{eqnarray*}
by Wick calculus in $(S)^*$, $Y(t)$ is the solution of the
following fractional SDE
\begin{eqnarray*}\left\{\begin{array}{rcl}
\displaystyle\frac{d}{dt}Y(t)&=&Y(t)\beta^{\prime}(t)+Y(t)\diamond
(\alpha
W_H(t))+Y(t)H\alpha^2t^{2H-1},\\
Y(0)&=&\exp(\beta(0)).\end{array}\right.
\end{eqnarray*}
Hence
$$
Y(t)=Y(0)+\int_0^tY(s)\beta^{\prime}(s)ds+\int_0^tY(s)\alpha
dB_H(s)+H\int_0^tY(s)\alpha^2s^{2H-1}ds,
$$
which means that
\begin{eqnarray}\label{Gitog2}
g(t,B_H(t))=g(0,0)+\int_0^t\displaystyle\frac{\partial g}{\partial
s }(s,B_H(s))ds+\int_0^t\displaystyle\frac{\partial g}{\partial x
}(s,B_H(s))dB_H(s)+H\int_0^t\displaystyle\frac{\partial^2
g}{\partial x^2 }(s,B_H(s))s^{2H-1}ds.
\end{eqnarray}
For $f(t,x)\in C^{1,2}(R\times R)$, we can find a sequence
$f_n(t,x)$ of a linear combinations of function $g(t,x)$ in
$(\ref{Gitog})$ such that $f_n(t,x), \displaystyle\frac{\partial
f_n(t,x) }{\partial t}, \frac{\partial f_n(t,x) }{\partial x},$
and $\displaystyle\frac{\partial^2 f_n(t,x) }{\partial x^2}$
pointwisely convergence to $f(t,x), \displaystyle\frac{\partial
f(t,x) }{\partial t}, \frac{\partial f(t,x) }{\partial x},$ and $
\displaystyle\frac{\partial^2 f(t,x) }{\partial x^2}$,
respectively, in $C^{1,2}(R\times R)$, then $f_n(t,x)$ satisfying
$(\ref{Gitog2})$ for all $n$. Since
\begin{eqnarray*}
&&\int_0^t\displaystyle\frac{\partial f_n}{\partial
x}(s,B_H(s))dB_H(s)\nonumber\\
&=&\int_0^t\displaystyle\frac{\partial f_n}{\partial
x}(s,B_H(s))\diamond W_H(s)d(s)\nonumber\\
&\longrightarrow &\int_0^t\displaystyle\frac{\partial f}{\partial
x}(s,B_H(s))\diamond W_H(s)d(s)\ \ \ \ \mbox{in } (S)^*,\ \mbox{as
} n\longrightarrow \infty.
\end{eqnarray*}
We prove the theorem.\ \ $\square$

\subsection{Fractional Differentiation}
Similar with the approach to differentiation in Aase et al.
\cite{Aase} (2000), Hu $\&$ $\O$ksendal \cite{Hu} (2003), and
Elliott $\&$ Van der Hoek \cite{Elliott} (2004), we define the
fractional differentiation of the function defined on the white
noise space $(S^{\prime}(R), S(R),\hat{E})$
\begin{definition}
Suppose $F: S^{\prime}(R)\longrightarrow R$ and suppose $\gamma\in
S^{\prime}(R)$. We say $F$ has a directional $M_H$-derivative in
the direction $\gamma$ if
$$
D_{\gamma}^{(H)}F(\omega):=\lim_{\varepsilon\longrightarrow
0}\displaystyle\frac{F(\omega+\varepsilon
M_H\gamma)-F(\omega)}{\varepsilon}
$$
exists in G-Hida space $(S)^*$, and $D_{\gamma}^{(H)}F(\omega)$ is
called the directional $M_H$-derivative of $F$ in the direction
$\gamma$.
\end{definition}
\begin{definition}
We say that $F:S^{\prime}(R)\longrightarrow R$ is
$M_H$-differentiable if there is a map
$$
\Psi: R\longrightarrow (S)^*
$$
such that $(M_H\Psi(t))\cdot (M_H\gamma)(t)$ is $(S)^*$ integrable
and
$$
D_{\gamma}^{(H)}F(\omega)=<F,\gamma>_{M_H},\ \ \mbox{for all }
\gamma\in L^2_{H}(R),
$$
where $<F,\gamma>_{M_H}:=\int_R(M_H\Psi(t))\cdot
(M_H\gamma)(t)dt$. We then define
$$
D_t^{(H)}F(\omega):=\displaystyle\frac{\partial ^{H}}{\partial
\omega }F(t,\omega)=\Psi(t,\omega)
$$
and we call $D_t^{(H)}F(\omega)$ the Malliavin derivative or
stochastic $M_H-$gradient of $F$ at $t$.
\end{definition}
\begin{definition}
Suppose $k\in \{1,2,\dots\}$ and for $n=0,1,2,\dots\
f_n\in\widehat{L}^2_{H}(R^n)$. We say that
$$
\Psi=\sum_{n=0}^{\infty}\int_{R^n}f_ndB^{\otimes n}_{H}(x)
$$
belongs to the space $\mathcal{G}_k(M_H)$ if
$$
\|\Psi\|^2_{\mathcal{G}_k}:=\sum_{n=0}^{\infty}n!\|M_H^nf_n\|^2_{L^2(R^n)}e^{2kn}<\infty.
$$
Define $\mathcal{G}(M_H)=\bigcap_k\mathcal{G}_k(M_H)$ and give
$\mathcal{G}(M_H)$ the projective topology.
\end{definition}
\begin{definition}
Suppose $q\in\{1,2,\dots\}$. We say the formal expansion
$$
Q=\sum_{n=0}^{\infty}\int_{R^n}g_ndB_{H}^{\otimes n}(x),
$$
with $g_n\in \widehat{L}^2_{H}(R^n),n=0,1,2,\dots,$ belongs to the
space $\mathcal{G}_{-q}(M_H)$ if
$$
\|Q\|_{\mathcal{G}_{-q}}^2=\sum_{n=0}^{\infty}n!\|M_H^ng_n\|_{L^2(R^n)}^2e^{-2qn}<\infty.
$$
\end{definition}
Define
$\mathcal{G}^*(M_H)=\bigcup_{q=1}^{\infty}\mathcal{G}_{-q}(M_H)$
and given $\mathcal{G}^*(M_H)$ the inductive topology. Then
$\mathcal{G}^*(M_H)$ is the dual space of $\mathcal{G}(M_H)$. For
$Q\in \mathcal{G}^*(M_H)$ and $\Psi\in \mathcal{G}(M_H)$ define
$$
<<Q,\Psi>>_{\mathcal{G}}=\sum_{n=0}^{\infty}n!<M_H^ng_n,M_H^nf_n>_{L^2(R^n)}.
$$
\begin{definition}\label{quasi}
Suppose
$$
Q=\sum_{n=0}^{\infty}\int_{R^n}g_n(s)dB_{H}^{\otimes n},
$$
is in $\mathcal{G}^*(M_H)$. We define the {\bf quasi-G-conditional
expectation} of $Q$ with respect to
$\mathcal{F}_t^{M_H}=\sigma\{B_{H}(s),0\leq s\leq t\}$ as
$$
\tilde{E}_{M_H}[Q|\mathcal{F}_{t}^{M_H}]:=\sum_{n=0}^{\infty}\int_{R^n}g_n(s)I_{n,(0,t)}(s)dB_{H}^{\otimes
n}(s),
$$
where $I_{n,(0,t)}(s)=I_{(0,t)}(s_1)\cdots I_{(0,t)}(s_n)$. We say
that $Q\in \mathcal{G}^*(M_H)$ is {\bf $\mathcal{F}_t^{M_H}$
measurable} if
$$
\tilde{E}_{M_H}[Q|\mathcal{F}_t^{M_H}]=Q
$$
\end{definition}
Note that, if $H\neq \frac{1}{2}$ the quasi-G-conditional
expectation
$\tilde{E}_{M_H}[B_{H}|\mathcal{F}_s^{M_H}]=B_{H}(s)\neq
\hat{E}[B_{H}|\mathcal{F}_s^{M_H}]$.
\begin{definition}
Suppose $F=\sum_{\alpha}c_{\alpha}H_{\alpha}\in
\mathcal{G}^*(M_H)$. Then the stochastic $M_H-$gradient of $F$ at
$t$ is defined by:
\begin{eqnarray*}
D_t^{H}F(\omega)&=&\sum_{\alpha}c_{\alpha}(\sum_{i}\alpha_i
H_{\alpha-\varepsilon_i}(\omega)e_i(t))\nonumber\\
&=&\sum_{\beta}(\sum_{i}c_{\beta+\varepsilon_i}(\beta_i+1)e_i(t))H_{\beta}(\omega).
\end{eqnarray*}
\end{definition}
From now on we denote $\nabla_t^HQ:=\tilde{E}_{M_H}[D_t^H Q
|\mathcal{F}_t^{M_H}]$.

\begin{theorem} {\bf (Fractional G-Clark-Ocone formula for polynomials)}
Suppose
$P(x)=\sum_{\alpha}c_{\alpha}x^{\alpha},x=(x_1,\dots,x_n)$, is a
polynomial. Consider $f_i\in L_{M_H}^2(R), 1\leq i\leq n$, and
$$
X_i^{(t)}(\omega)=\int_0^tf_idB_{H}(\omega)=<M_H(I_{(0,t)}f_i),\omega>.
$$
With $X^{t}=(X_1^{(t)},\dots,X_n^{(t)})$ suppose that
$$
F=P^{\diamond}(X^{(t)}).
$$
Then
\begin{eqnarray}
F=\hat{E}(F)+\int_0^T\tilde{E}_{M_H}[D_t^{H}F|\mathcal{F}_t^{M_H}]dB_{H}(t).
\end{eqnarray}
\end{theorem}
{\bf Proof.} From the Definition $\ref{quasi}$ of the
quasi-G-conditional expectation, we see that $F$ is
$\mathcal{F}_T^{M_H}$ measurable, so
\begin{eqnarray*}
F(\xi)&=&\tilde{E}_{M_H}[F|\mathcal{F}_{T}^{M_H}]\\
&=&\sum_{\alpha}c_{\alpha}\tilde{E}_{M_H}[X|\mathcal{F}_{T}^{M_H}]^{\diamond
\alpha } \\
&=&P^{\diamond}(X^{T}).
\end{eqnarray*}
Thus we have
\begin{eqnarray*}
\int_0^T\tilde{E}_{M_H}[F|\mathcal{F}_{T}^{M_H}]dB_{H}(t)&=&\int_0^T\tilde{E}_{M_H}[\sum_{i=1}^{n}(\frac{\partial
P }{\partial x_i})^{\diamond}(X)f_i(t)|\mathcal{F}_{T}^{M_H}]dB_{H}(t)\\
&=&\int_0^T\frac{d}{dt}P^{\diamond}(X^{(t)})dt\\
&=&P^{\diamond}(X^{(T)})-P^{\diamond}(X^{(0)})\\
&=&F-\hat{E}[F].\ \ \ \ \ \ \ \ \ \ \square
\end{eqnarray*}

Using the similar argument in Aase et al. \cite{Aase} (2000), Hu
$\&$ $\O$ksendal \cite{Hu} (2003), and Elliott $\&$ Van der Hoek
\cite{Elliott} (2004), we can establish the following fractional
G-Clark-Ocone Theorem

\begin{theorem} {\bf (Fractional G-Clark-Ocone Theorem)}
(a) Suppose $Q\in \mathcal{G}^*(M_H)$ is $\mathcal{F}_T^{M_H}$
measurable. Then $D_t^{H}Q\in \mathcal{G}^*(M_H)$ and
$\tilde{E}_{M_H}[D_t^{H}Q|\mathcal{F}_T^{M_H}]\in
\mathcal{G}^*(M_H)$ for almost all $t$.
$\tilde{E}_{M_H}[D_t^{H}Q|\mathcal{F}_T^{M_H}]\diamond W_H(t)$ is
integrable in $S^*$ and
\begin{eqnarray}
Q=\hat{E}[Q]+\int_0^T\tilde{E}_{M_H}[D_t^{H}Q|\mathcal{F}_T^{M_H}]\diamond
W_{H}(t)dt.
\end{eqnarray}
(b) Suppose $Q\in L^2_{M_H}$ is $\mathcal{F}_T^{M_H}$ measurable.
Then $\tilde{E}_{M_H}[D_t^{H}Q|\mathcal{F}_T^{M_H}](\omega)\in
L^{1,2}_{M_H}(R)$ (the definition of $L^{1,2}_{M_H}(R)$ is the
analogue in Elliott $\&$ Hoek \cite{Elliott}) and
\begin{eqnarray}
Q=\hat{E}[Q]+\int_0^T\tilde{E}_{M_H}[D_t^{H}Q|\mathcal{F}_T^{M_H}]dB_{H}(t).
\end{eqnarray}
\end{theorem}

\section{G-Girsanov's Theorem}
In this section we will construct G-Girsanov's Theorem, a analogue
in a special situation was proposed in \cite{Chen} defined from
PDE and \cite{Hu} in a view of SDE. For $\phi\in L^2(R)$, setting
\begin{eqnarray}
B^G(t):=B(t)-\int_0^t\phi(s)ds,
\end{eqnarray}
where $B(t)$ is a G-Brownian motion in sublinear space $(\Omega,
\mathcal{H},\mathcal{F},\hat{E})$, $\mathcal{F}_t=\sigma\{B(s),
s\leq t \}$. We will construct a time consistent G expectation
$E^G$, and transfer the sublinear expectation space $(\Omega,
\mathcal{H},\hat{E})$ to the sublinear expectation space
$(\Omega,\mathcal{H},E^G)$ such that $B^G(t)$ is a G-Brownian
motion on $(\Omega,\mathcal{H},E^G)$.

Define a sublinear function $G(\cdot)$ as follows
\begin{eqnarray}\label{eq415}
G(\alpha)=\displaystyle\frac{1}{2}(\overline{\sigma}^2\alpha^+-\underline{\sigma}^2\alpha^-),&\forall
\alpha\in R.
\end{eqnarray}
For given $\varphi\in C_{b,lip}(R)$, we denote $u(t,x)$ as the
viscosity solution of the following G-heat equation
\begin{eqnarray}\label{eq416}
\partial_tu-G(\partial_{xx}u)=0,&(t,x)\in(0,\infty)\times R,\\
u(0,x)=\varphi(x).&\nonumber
\end{eqnarray}
\begin{remark}
The G-heat equation ($\ref{eq416}$) is a special kind of
Hamilton-Jacobi-Bellman equation, also the Barenblatt equation
except the case $\underline{\sigma} = 0$ (see \cite{Baren1} and
\cite{Baren2}). The existence and uniqueness of ($\ref{eq416}$) in
the sense of viscosity solution can be found in, for example
\cite{Fleming}, \cite{Lions}, and \cite{PengA} for
$C^{1,2}$-solution if $\underline{\sigma} > 0$.
\end{remark}
The stochastic path information of $B^G(t)$ up to $t$ is the same
as $B(t)$, without loss of generality we still denote
$\mathcal{F}$ as the path information of $B^G(t)$ up to $t$.
Consider the process $B^G(t)$, we define
 $E^G[\cdot]: \mathcal{H}\longrightarrow R$ as
\begin{eqnarray*}
E^G[\varphi(B^G(t)]=u(t,0), t\in (0,+\infty)
\end{eqnarray*}
and for each $t,s\ge 0$ and $0< t_1< \cdots<t_N\leq t$
\begin{eqnarray*}
E^G[\varphi(B^G(t_1),\cdots,B^G(t_N),B^G(t+s)-B^G(t)]:=E^G[\psi(B^G(t_1),\cdots,B^G(t_N))]
\end{eqnarray*}
where $\psi(x_1,\cdots,x_N)=E^G[\varphi(x_1,\cdots,x_N,B^G(s)]$.

For $0<t_1<t_2<\cdots<t_{i}<t_{i+1}<\cdots<t_N<+\infty$, we define
G conditional expectation with respect to $\mathcal{F}_{t_i}$ as
\begin{eqnarray*}
&&E^G[\varphi(B^G(t_1),B^G(t_2)-B^G(t_1),\cdots,B^G(t_{i+1})-B^G(t_{i}),\cdots,
B^G(t_N)-B^G(t_{N-1})|\mathcal{F}_{t_i}]\\
&:=&\psi(B^G(t_1),B^G(t_2)-B^G(t_1),\cdots,
B^G(t_{i})-B^G(t_{i-1})),
\end{eqnarray*}
where
$\psi(x_1,\cdots,x_i)=E^G[\varphi(x_1,\cdots,x_i,B^G(t_{i+1})-B^G(t_{i}),
\cdots,B^G(t_{t_N})-B^G(t_{N-1})$.

By the comparison theorem of the G-heat equation ($\ref{eq416}$)
and the sublinear property of the function G($\cdot$), we
consistently define a sublinear expectation $E^G$ on
$\mathcal{H}$. Under sublinear expectation $E^G$ we define above
$B^G(t)$ is a G-Brownian motion and $B^G(t)$ is
$N(\{0\},[\underline{\sigma}^2t,\overline{\sigma}^2t])$
distributed. We call $E^G[\cdot]$ as G-expectation on
$(\Omega,\mathcal{H})$. Without loss generality. we denote
$B^G(t))$ as two-sided G-Brownian motion.

With a similar argument used in Section 2, we denote the Banach
space $(\hat{H}_p,\|\cdot\|_p)$ as $L_G^p(\Omega,\hat{H}_p,E^G)$,
and $B^G(t)$ is a G-Brownian motion in
$L_G^p(\Omega,\hat{H}_p,E^G)$, consequently,
\begin{eqnarray*}
B_H^G(t):&=&\int_RM_HI_{(0,t)}(s)dB^G(s)\\
&=&\int_RM_HI_{(0,t)}(s)dB(s)-\int_RM_HI_{(0,t)}(s)\phi(s)ds
\end{eqnarray*}
is a fGBm under G-expectation $E^G$.

To eliminate a drift in the fGBm we need to solve equations of the
form
$$
\int_RM_HI_{(0,t)}(s)\phi(s)ds=g(t).
$$
That is
$$
\int_0^tM_H\phi(s)ds=g(t).
$$
Consequently, with $(M_H\phi)(t)=g^{\prime}(t)$
$$
\phi(t)=(M_H^{-1}g^{\prime})(t).
$$
If $g^{\prime}(t)=A$ on $[0,T]$
\begin{eqnarray*}
\phi(t)&=&AM_{1-H}I_{(0,T)}(t)\\
&=&\displaystyle\frac{A(\sin(\pi(1-H))\Gamma(3-2H))^{\frac{1}{2}}}{2\Gamma
(\frac{1}{2}-H)\cos{(\frac{\pi}{2}(\frac{1}{2}-H))}}\left[\displaystyle\frac{T-t}{|T-t|^{\frac{1}{2}+H}}+\frac{t}{|t|^{\frac{1}{2}+H}}\right],
\end{eqnarray*}
and for $0\leq t\leq T$
\begin{eqnarray*}
\phi(t)&=&\displaystyle\frac{A(\sin(\pi(1-H))\Gamma(3-2H))^{\frac{1}{2}}}{2\Gamma
(\frac{1}{2}-H)\cos{(\frac{\pi}{2}(\frac{1}{2}-H))}}\left[(T-t)^{\frac{1}{2}-H}+(t)^{\frac{1}{2}-H}\right].
\end{eqnarray*}
\begin{theorem}(G-Girsanov Theorem) Assume that in the sublinear expectation space $(\Omega, \mathcal{H},\hat{E})$ the stochastic process $B_H(t)$ is a fGBm
with Hurst index $H\in (0,1)$, then there exists G-expectation
$E^G$, such that, in the G-expectation space $(\Omega,
\mathcal{H},E^G)$
\begin{eqnarray*}
B^G(t):=B(t)-\int_0^t\phi(s)ds,
\end{eqnarray*}
is a G-Brownian motion with notation $B(t)=B_{\frac{1}{2}}(t)$,
and
\begin{eqnarray*}
B_H^G(t):&=&\int_RM_HI_{(0,t)}(s)dB(s)-\int_RM_HI_{(0,t)}(s)\phi(s)ds
\end{eqnarray*}
is a fGBm under G-expectation $E^G$ with drift
$\int_RM_HI_{(0,t)}(s)\phi(s)ds=g(t)$, and
$\phi(t)=(M_{1-H}g^{\prime})(t)$.
\end{theorem}

\section{Financial Application}

We start from a family of probability space with equivalent
probability measure
$\{(S^{\prime}(R),S(R),\mathcal{F},P^{\theta}):\theta\in \Theta\}$
where $P^{\theta}$ is an induced probability by
$\int_0^t\sigma^{\theta}_tdB^0(t)$ with $B^0(t)$ be a standard
Brownian motion in reference space $((S^{\prime}(R),S(R),,F,P_0)$,
$\mathcal{F}$ is the augment filter constructed from the Brownian
motion $B^0(t)$, and $(\sigma^{\theta}_t)_{t\ge 0}$ are unknown
processes parameterized by $\theta\in \Theta$, and $\Theta$ is a
nonempty convex set. Here $\sigma^{\theta}_t$ means the
uncertainty of the volatility of the process
$\sigma^{\theta}_tB^0(t)$, we assume that

\begin{assum} $(H_1)$ Assume that $(\sigma^{\theta}_t)_{t\ge
0}$ is an adapt process for $\mathcal{F}_t$, and satisfying
$$
\sigma^{\theta}_t\in [\underline{\sigma},\overline{\sigma}],
\mbox{for all }\theta\in \Theta.
$$
\end{assum}
We set
$$
\widehat{E}[X]=\sup_{\theta\in \Theta}E_{\theta}[X],
$$
where $E_{\theta}$ is the correspond linear expectation of
probability $P^{\theta}$. Thus
$((S^{\prime}(R),S(R),\mathcal{F},\widehat{E})$ is a sublinear
expectation space, Denis, Hu and Peng \cite{Denis} prove that
$\sigma^{\theta}_tB^0(t)$ is a $G$-Brownian motion in sublinear
expectation space $((S^{\prime}(R),S(R),\mathcal{F},\widehat{E})$,
we denote $B_{\frac{1}{2}}:=\sigma^{\theta}_tB^0(t)$. Set
$B_H^0(t)$ is fBm with Hurst index $H\in (0,1)$ in
$((S^{\prime}(R),S(R),,\mathcal{F},P_0)$, then it is easy to prove
that $\sigma^{\theta}_tB_H^0(t)$ is a fGBm with Hurst index $H$ in
$(\Omega,\mathcal{F},\widehat{E})$, we denote it as $B_H(t)$.

We consider an incompleted financial market which contains a bond
$P(t)$ with
\begin{eqnarray}
dP(t)&=&rP(t)dt,\ \ 0\leq t\leq T,\\
 P(0)&=&1,\nonumber
\end{eqnarray}
and a stock whose price $S(t)$ with uncertain volatility and
satisfying the following SDE driven by a fGBm:
\begin{eqnarray}\label{stock1}
dS(t)&=&S(t)\diamond[\mu dt+d B_H(t)]\nonumber\\
&=&S(t)\diamond[\mu +W_H(t)]dt, \ \ 0\leq t\leq T,\\
S(0)&=&x.\nonumber
\end{eqnarray}
which equivalent to a family of $SDE$ with $\theta\in \Theta$
\begin{eqnarray}\label{stock2}
dS(t)&=&S(t)\diamond[\mu dt+\sigma_t^{\theta}d B^0_H(t)]\nonumber\\
&=&S(t)\diamond[\mu +\sigma_t^{\theta}W^0_H(t)]dt, \ \ 0\leq t\leq T,\\
S(0)&=&x,\nonumber
\end{eqnarray}
where $W_H^0(t)$ is fractional noise with respect to fBm
$B_H^0(t)$.

By G-Girsanov's Theorem, there exists G-expectation $E^G[\cdot]$
and G-Brownian motion $B^{G}(t)$ in the sublinear expectation
space $\mathcal{L}_G^p(S^{\prime}(R),S(R),E^G)\ \ (p\geq 1)$, such
that
\begin{eqnarray}
B^G(t)=(\mu-r)t+B(t).
\end{eqnarray}
Under $E^G$
\begin{eqnarray}
B^G_H(t):=(\mu-r)t+B_H(t),\ \ (\mbox{or } W^G_H:=(\mu-r)+W_H(t)),
\end{eqnarray}
is a fGBm with Hust index $H$, and
\begin{eqnarray*}
\phi(t)=(r-\mu)M_{1-H}I_{(0,T)}(t).
\end{eqnarray*}

Then we have
\begin{eqnarray}\label{stock3}
dS(t)&=&S(t)\diamond[r dt+d B^G_H(t)]\nonumber\\
&=&S(t)\diamond[r +W^G_H(t)]dt, \ \ 0\leq t\leq T.
\end{eqnarray}

By fractional G-It\^o formula
\begin{eqnarray*}
S(t)=x\exp{(r t+B^G_H(t)-\frac{1}{2}t^{2H})}
\end{eqnarray*}
is the solution to the SDE $(\ref{stock3})$

Suppose that a portfolio $\pi(t)=(u(t),v(t))$ be a pair of
$\mathcal{F}_t^H$ adapted processes, where $\mathcal{F}_t^H$ is
the augment filter of fGBm $B^G_H(t)$. The corresponding wealth
process is
\begin{eqnarray*}
W^{\pi}(t)=u(t)P(t)+v(t)S(t),
\end{eqnarray*}
where $\pi$ is called admissible if $W^{\theta}(t)$ is bounded
below for all $t\in [0,T]$ and $\pi$ is self-financing if
\begin{eqnarray*}
dW^{\pi}(t)&=&u(t)dP(t)+v(t)S(t)\diamond [r
dt+dB^G_H(t)]\nonumber\\
&=&rW^{\pi}(t)dt+v(t)S(t)\diamond W^G_H(s)ds.
\end{eqnarray*}
\begin{eqnarray*}
e^{-rt}W^{\pi}(t)=W^{\pi}(0)+\int_0^te^{-rs}v(s)S(s)\diamond
W^G_H(s)ds.
\end{eqnarray*}

Suppose there is a negative and $\mathcal{F}_T^H-$ measurable
contingent claim $\xi\in \mathcal{L}_G^2(S^{\prime}(R),S(R),E^G)$
with maturity $T>0$ in the market,
applying our fractional G-Clark-Ocone Theorem to
$e^{-rt}\xi(\omega)$, we have
\begin{eqnarray}\label{GV}
e^{-rT}\xi=E^G[e^{-rT}\xi]+\int_0^T\tilde{E}_{M_H}[e^{-rT}D_t^H\xi|\mathcal{F}_t^H]\diamond
W_H^G(t)dt,
\end{eqnarray}
where $\tilde{E}_{M_H}$ is the quasi-G-conditional expectation
defined in Definition $\ref{quasi}$, $D_t^H$ is the G-fractional
Malliavin differential operator.

On the Probability framework, by Girsanov's Theorem given in
\cite{Elliott}
\begin{eqnarray}\label{transform2}
\widehat{B}_H(t):=\displaystyle\frac{\mu-r}{\sigma^{\theta}_t}t+B^0_H(t),\
\ (\mbox{or }
\widehat{W}_H(t):=\displaystyle\frac{\mu-r}{\sigma^{\theta}_t}t+W^0_H(t))
\end{eqnarray}
is a fBm with respect to the measure $\widehat{P}^{\theta}$ by
$$
\displaystyle\frac{d\widehat{P}^{\theta}}{P^{\theta}}(\omega)=\exp[<\phi^{\theta},\omega>-\frac{1}{2}\|\phi^{\theta}\|^2]
$$
where $\phi^{\theta}(t)=\phi(t)/\sigma_t^{\theta}$. We can rewrite
the equation $(\ref{transform2})$ as
\begin{eqnarray*}
\sigma^{\theta}_t\widehat{B}_H(t):=(\mu-r)t+\sigma^{\theta}_tB^0_H(t),\
\ (\mbox{or }
\sigma^{\theta}_t\widehat{W}_H(t):=(\mu-r)t+\sigma^{\theta}_tW^0_H(t)).
\end{eqnarray*}
Define
$$
\mathcal{E}[X]:=\sup_{\theta\in
\Theta}E_{\widehat{P}^{\theta}}[X], \ \ X\in\mathcal{H},
$$
where $E_{\widehat{P}^{\theta}}[\cdot]$ is the linear expectation
corresponding to the probability measure $\widehat{P}^{\theta}$.
Denis, Hu and Peng \cite{Denis} (2010) prove that under the
sublinear expectation $\mathcal{E}[\cdot]$,
$\sigma_t^{\theta}\widehat{B}_{1/2}(t)$ is a G-Brownian motion on
the sublinear expectation space $(S^{\prime}(R),S(R),E^G)$. Notice
that
$$
B^{G}(t)\stackrel{d}{=}\sigma_t^{\theta}\widehat{B}_{1/2}(t)
$$
and $B^{G}(t)$ is G-Brownian motion on the sublinear expectation
space $(S^{\prime}(R),S(R),E^G)$, thus
\begin{eqnarray*}
E^G[\cdot]&=&\mathcal{E}[\cdot],
\end{eqnarray*}
i.e.,
\begin{eqnarray}\label{EG_SUP}
 E^G[X]&=&\sup_{\theta\in
\Theta}E_{\widehat{P}^{\theta}}[X], \ \ X\in\mathcal{H}.
\end{eqnarray}
It is easy to proof that $\sigma_t^{\theta}\widehat{B}_H(t)$ is
fGBm on the sublinear expectation space
$(S^{\prime}(R),S(R),E^G)$.

By using the Girsnov's transform $(\ref{transform2})$, the stock
price process $(\ref{stock2})$ can be rewritten as
\begin{eqnarray*}\label{stock}
dS(t)&=&S(t)\diamond[r dt+\sigma_t^{\theta}d \widehat{B}_{H}(t)]\nonumber\\
&=&S(t)\diamond[r +\sigma_t^{\theta}\widehat{W}_H(t)]dt, \ \ 0\leq t\leq T,\\
S(0)&=&x.\nonumber
\end{eqnarray*}
Elliott and Hoek \cite{Elliott}, and Hu and $\O$ksendal \cite{Hu}
prove that
$$
E_{\widehat{P}^{\theta}}[e^{-rT}\xi]
$$
is the price of the claim and
$$
v(t)=S(t)^{-1}e^{-r(T-t)}(\sigma^{\theta}_t)^{-1}\widetilde{E}_H^{\theta}[\tilde{D}_t^H\xi|\mathcal{F}_t^H]
$$
determines the portfolio, such that
\begin{eqnarray}\label{P-GV}
e^{-rT}\xi=E_{\widehat{P}^{\theta}}[e^{-rT}\xi]+\int_0^T\widetilde{E}_H^{\theta}[e^{-rT}D_t^H\xi|\mathcal{F}_t^H]\diamond
W_H^G(t)dt,
\end{eqnarray}

where $\widetilde{E}_H^{\theta}$ is the quasi-conditional
expectation, and $\tilde{D}_t^H$ is the fractional Malliavin
differential operator defined in \cite{Elliott}.

From $(\ref{EG_SUP})$, we have
\begin{eqnarray}
E^G[e^{-rT}\xi]=\sup_{\theta\in\Theta}E_{\widehat{P}^{\theta}}[e^{-rT}\xi],
\end{eqnarray}
thus, notice $(\ref{GV})$, $E^G[e^{-rT}\xi]$ is the bid price of
the claim $\xi$ with
$$
v(t)=S(t)^{-1}e^{-r(T-t)}\tilde{E}_{M_H}[D_t^H\xi|F_t^H]
$$
determines the portfolio in the super hedging. Similarly, we can
derive that
\begin{eqnarray}
-E^G[-e^{-rT}\xi]=\inf_{\theta\in\Theta}E_{\widehat{P}^{\theta}}[e^{-rT}\xi],
\end{eqnarray}
is the ask price of the claim.

\end{document}